# The William Kruskal Legacy: 1919–2005[1]

**Stephen E. Fienberg, Stephen M. Stigler and Judith M. Tanur**

*Abstract.* William Kruskal (Bill) was a distinguished statistician who spent virtually his entire professional career at the University of Chicago, and who had a lasting impact on the Institute of Mathematical Statistics and on the field of statistics more broadly, as well as on many who came in contact with him. Bill passed away last April following an extended illness, and on May 19, 2005, the University of Chicago held a memorial service at which several of Bill's colleagues and collaborators spoke along with members of his family and other friends. This biography and the accompanying commentaries derive in part from brief presentations on that occasion, along with recollections and input from several others.

Bill was known personally to most of an older generation of statisticians as an editor and as an intellectual and professional leader. In 1994, *Statistical Science* published an interview by Sandy Zabell (Vol. 9, 285–303) in which Bill looked back on selected events in his professional life. One of the purposes of the present biography and accompanying commentaries is to reintroduce him to old friends and to introduce him for the first time to new generations of statisticians who never had an opportunity to interact with him and to fall under his influence.

*Stephen E. Fienberg is Maurice Falk University Professor of Statistics and Social Science in the Department of Statistics, the Center for Automated Learning and Discovery, and Cylab, all at Carnegie Mellon University, Pittsburgh, Pennsylvania 15213-3890, USA e-mail: fienberg@stat.cmu.edu. Stephen M. Stigler is Ernest DeWitt Burton Distinguished Service Professor and Chairman, Department of Statistics and the College, and member of the Committee on Conceptual and Historical Studies of Science, University of Chicago, Chicago, Illinois 60637, USA e-mail: stigler@galton.uchicago.edu. Judith M. Tanur is Distinguished Teaching Professor Emerita, Department of Sociology, State University of New York, Stony Brook, New York 11794-4356, USA e-mail: jtanur@notes.cc.sunysb.edu.*

## 1. A BRIEF BIOGRAPHY

William Henry Kruskal was born in New York City on October 10, 1919, the oldest of three boys and two girls. His father, Joseph Kruskal, owned Kruskal & Kruskal, which was for many decades the nation's largest wholesale fur business. Bill's mother, Lillian Vorhaus Kruskal, later became famous as Lillian Oppenheimer, founder of what is today called Origami USA, co-authoring books on origami and making numerous television appearances to promote the art of paperfolding. (She also taught origami at summer camp to Judy Tanur's children.)

The three Kruskal brothers all went on to research careers in related fields. "Bill, Martin and I all started as mathematicians, but Bill moved completely into statistics, I moved partially into statistics and Martin moved partially into physics," noted Bill's brother Joseph Kruskal, Jr., widely known for "Kruskal's theorem" in computer science and his work on multidimensional scaling, and now retired from Bell Laboratories. Bill's other brother, Martin Kruskal, a professor emeritus of mathematics at Princeton University and now at Rutgers University, in 1993 received the National Medal of Science, the nation's highest award for scientific achievement.

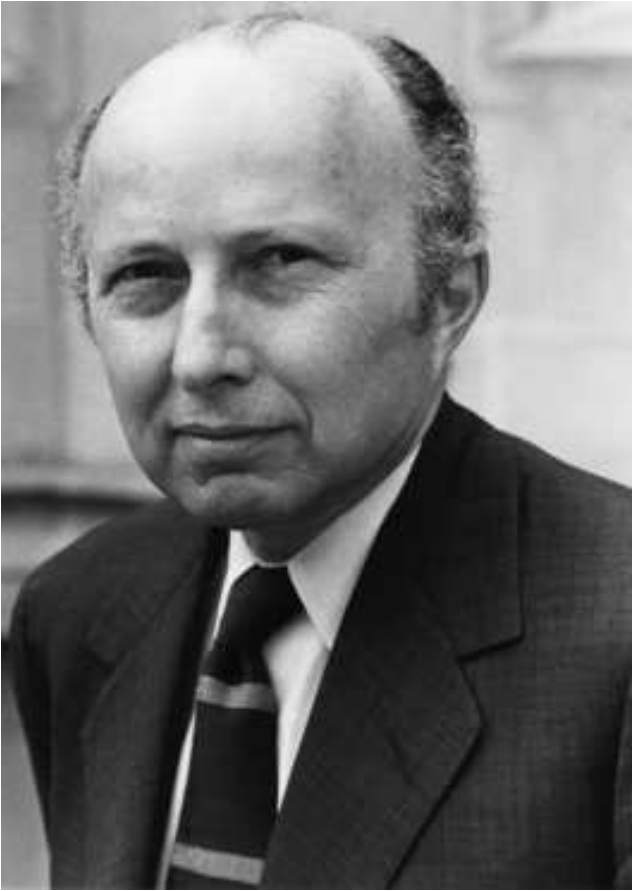

Fig. 1.   *William H. Kruskal: 1919–2005.*

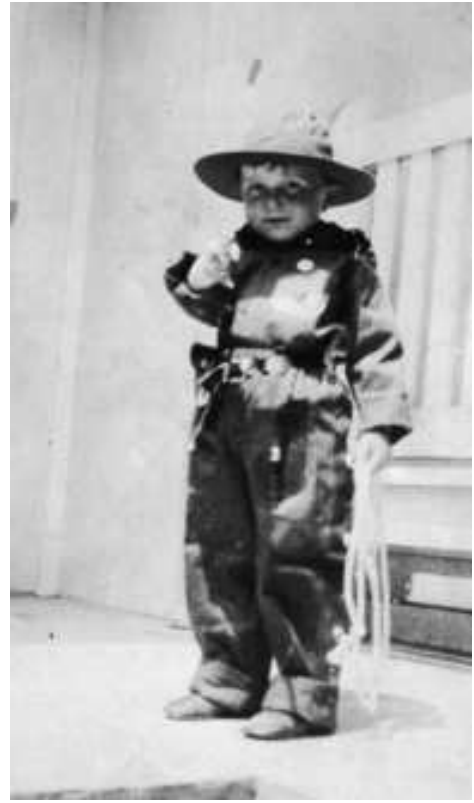

Fig. 2.   *"Bill, save this picture. You were so very proud of your cowboy suit with the gun."*

Bill first attended Antioch College and then transferred to Harvard University, receiving his bachelor's degree in mathematics and philosophy summa cum laude in 1940. He then received his master's degree in mathematics from Harvard in 1941 and his Ph.D. in mathematical statistics from Columbia University in 1955.

Bill was a mathematician at the U.S. Naval Proving Ground in Dahlgren, Virginia, from 1941 to 1946, and worked for Kruskal & Kruskal from 1946 to 1948. He was a lecturer in mathematics at Columbia

University in 1949–1950. He joined the University of Chicago faculty as an instructor in statistics in 1950 and progressed through the ranks to full professor. Along the way, he took brief appointments as a visiting professor at the University of California, Berkeley in 1955–1956, and at Harvard University in the summer of 1959. Bill was named the Ernest DeWitt

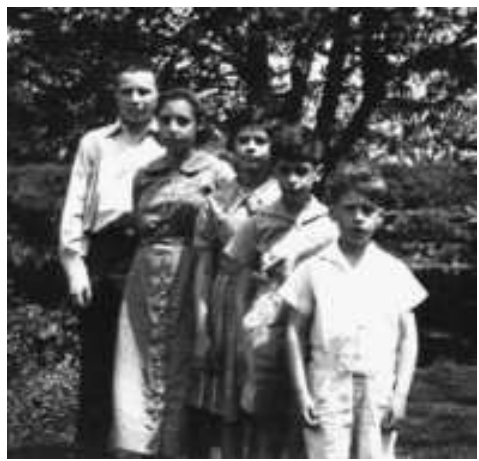

Fig. 3.   *The Kruskal siblings: Bill, Molly, Rosaly, Martin David and Joe.*

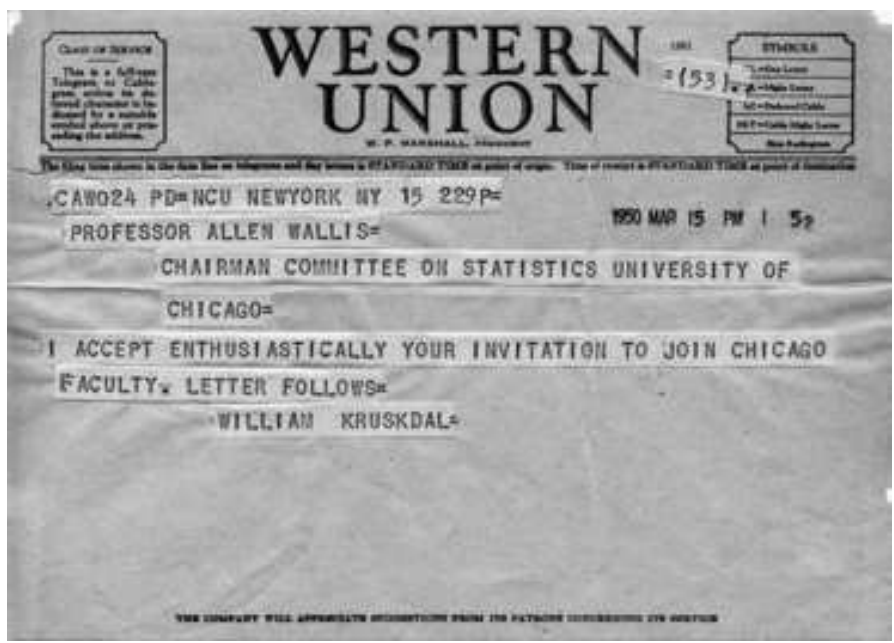

Fig. 4. *Western Union telegram from Kruskal to Wallis, accepting the initial appointment at the University of Chicago.*

Burton Distinguished Service Professor in Statistics in 1973. His wide-ranging interdisciplinary research and leadership reached across two academic divisions and a professional school at the University of Chicago over the decades. He was a founding faculty member of the Department of Statistics in the Physical Sciences Division and served as department Chairman from 1966 to 1973, playing a vital role in building the new Department of Statistics and establishing an unusually effective collegiality within the department. He further served the University as dean of the Social Sciences Division from 1974 to 1984, and as dean pro tempore of the newly established Irving B. Harris Graduate School of Public Policy Studies from 1988 to 1989. He retired as professor emeritus in 1990. In separate commentaries, Stephen Stigler comments on Bill's role within the department and Norman Bradburn focuses on Bill's role as a citizen of the university and of the broader statistical and social science community.

A reference letter dated February 16, 1950 from Abraham Wald to W. Allen Wallis, founding chairman of the then Committee on Statistics (which later became the Department of Statistics), included the following comments:

> Very gifted and intelligent. Belongs in the upper 5 percent of our graduate students. Has a very good mathematical background

(master's degree in mathematics from Harvard). Very well equipped and capable for theoretical work in statistics, but probably more interested in applications. He was at the Naval Proving Ground at Dahlgren, Virginia during the war and, therefore, had experience in the application of mathematics to practical problems. He is mature, conscientious and industrious and has a pleasant personality. He has made a good beginning on his Ph.D. thesis work. Doubtful whether he can finish it by the end of this academic year. Age: 30 years.

[Wald died December 13, 1950.] Bill was appointed originally as instructor for a one-year period, October 1, 1950 to September 30, 1951. The formal offer letter from Wallis (March 6, 1950) makes no reference to a Ph.D. degree. Bill's acceptance came just over a week later by telegram, reproduced here as Figure 4.

Bill was appointed an assistant professor for a three-year term as of October 1, 1951 and reappointed on December 14, 1953 for another three-year term. There is no mention of his Ph.D. degree in the proposal letter from Wallis or in the Dean's appointment letter. Bill was a perfectionist and it was not until 1955 that he submitted the final version of his thesis and received his Ph.D. (Departments were then clearly more forgiving of such matters than they are today.) Bill went on leave to Berkeley for the 1955–1956 academic year and Wallis initiated his promotion to associate professor with indefinite tenure that fall, informing the dean of the threat that Bill might stay at Berkeley. Bill's promotion became official in December 1956 and took effect on October 1, 1957.

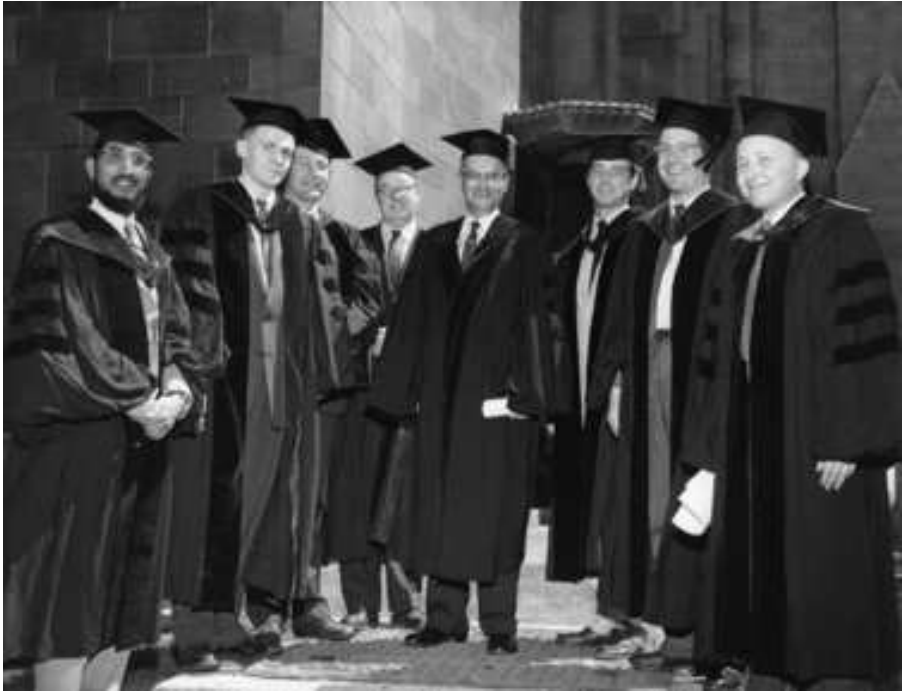

FIG. 5. *Bill and other statistics faculty with Jerzy Neyman in 1959 on the occasion of Neyman's receipt of an honorary degree from the University of Chicago.*

Bill became editor of *The Annals of Mathematical Statistics* in 1958 and he served in that capacity until 1961. This, however, was only the beginning of Bill's career as an editor, an activity to which he returned repeatedly throughout his career. He was president of the Institute of Mathematical Statistics in 1970–1971 and of the American Statistical Association in 1982.

Along with senior colleague W. Allen Wallis, Bill devised the Kruskal–Wallis test, a now ubiquitous rank test for one-way analyses of variance, found today under that name as part of every major statistical computing package. Their original paper appeared in the *Journal of the American Statistical Association* (*JASA*) in 1952, and it and the test remain widely cited today (see the discussion in Stigler's accompanying commentary).

With another Chicago colleague, Leo Goodman, Bill co-authored a series of four classic papers that, returning to first principles, brought new insight to measuring the association between pairs of qualitative attributes. These appeared over a 18-year period in *JASA*, beginning in 1954, and were later republished in book form in 1979. Goodman describes some of their work and interactions in an accompanying commentary. Bill also wrote a separately authored piece in *JASA* on ordinal measures of association which dovetailed with the Goodman–Kruskal work.

Throughout this period, Bill worked on and off on his coordinate-free approach to linear statistical models. Although he published only three brief papers on the topic, he also developed what became a legendary set of unpublished class notes elaborating on the ideas. In this work Bill freed statisticians from relying on vision-clouding coordinate frames of reference. Morris Eaton, who was a junior colleague to Bill in the late 1960s and early 1970s, describes the impact of this work in an accompanying commentary.

Bill co-authored yet another series of landmark papers with Frederick Mosteller in the 1980s on representative sampling. These papers trace the history of the notion of probability sampling and describe how it evolved from the "loose" and not-well-defined notion of "representativeness." Bill had a fascination with the origins of statistical terminology and ideas. This fascination also is reflected in his work with Steve Stigler on the appearance and uses of "normal" in statistics (normal equations, normal distribution, with senses of both the usual and the ideal), and in Bill's longstanding investigation of the term and idea of "relative importance."

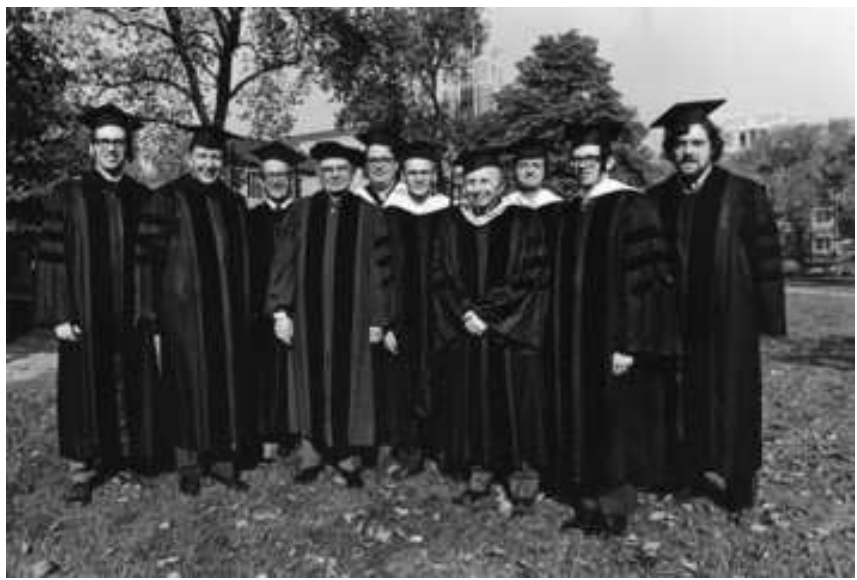

FIG. 6. *Bill and other statistics faculty with Frederick Mosteller in 1973 on the occasion of Mosteller's receipt of an honorary degree from the University of Chicago.*

President Richard Nixon appointed Bill to his President's Commission on Federal Statistics in 1970. The 15-member Commission conducted a comprehensive review of the data-gathering and compilation and use of statistics by the federal government, the first such review that had taken place in 20 years. The subsequent creation of a permanent committee to address the kinds of problems the President's Commission had laid out (the National Research Council's Committee on National Statistics) was predominantly due to Bill's efforts, and he became its first chairman, serving from 1971 to 1978. The Committee was charged with evaluating statistical issues for the U.S. government, including citizens' attitudes and behavior toward the census. Margaret Martin in an accompanying commentary describes her interactions with Bill as they launched the Committee.

Bill's academic interests were encyclopedic. Indeed, he put these interests to use as the associate editor for statistics of the *International Encyclopedia of the Social Sciences* from 1962 to 1968, and co-editor (with Judith Tanur) of the *International Encyclopedia of Statistics* in 1978. His role as editor is described by Tanur in her accompanying commentary and also by Stephen Fienberg.

When a statistical issue in an unfamiliar setting caught his eye, Bill would frequently begin a file that would grow over time, as he added the products of his library work, correspondence he would initiate with experts in other fields, and clippings from his voracious reading of newspapers, scholarly magazines and professional journals. The topics would range from cloud seeding to casualty counts in the Vietnam War to the Dreyfus Affair. These files and the careful references to the sources he had encountered added immeasurably to the depth of scholarship of all entries in the encyclopedias he edited. Even those contributing authors who in fact never read beyond narrow specialties acquired through Bill's uncredited editorial intervention the appearance of widely read scientific generalists. As several of the commentaries note, Bill's mailing of clippings became a source of amusement, stimulation and an occasional prod to action among his varied colleagues, friends and acquaintances.

Bill received many honors over the course of his career. In 1970–1971 he was a Fellow at the Center for Advanced Study in the Behavioral Sciences and NSF Senior Postdoctoral Fellow, and he was a Fellow of the John Simon Guggenheim Memorial Foundation in 1979–1980. He was elected a Fellow of the Institute of Mathematical Statistics, the American Statistical Association, the American Association for the Advancement of Science and the American Academy of Arts and Sciences, and he was an honorary Fellow of the Royal Statistical Society. In 1978, he served as the Committee of Presidents of Statistical Societies Fisher Lecturer and received the Samuel S. Wilks Memorial Medal from the American Statistical Association. In 1992 the American Statistical Association recognized Bill's long running fundamental contributions by giving him its Founders Award.

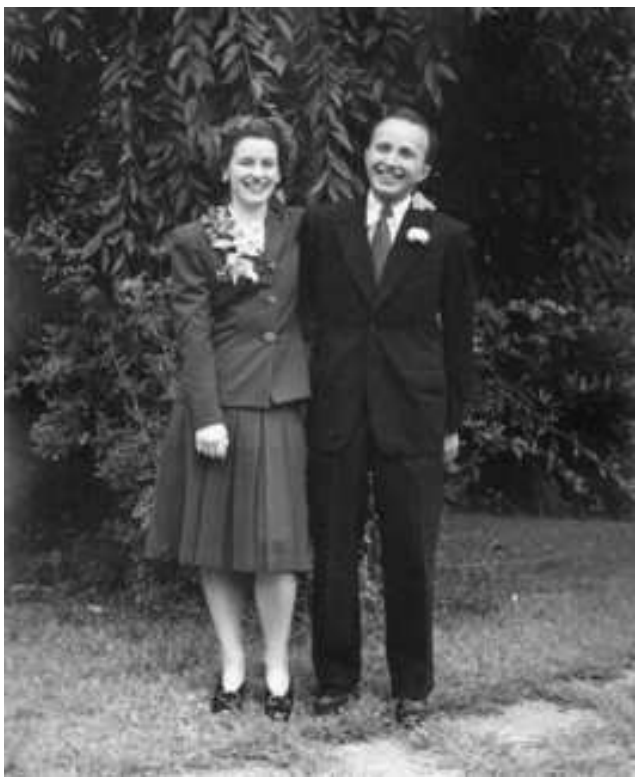

Fig. 7. *Bill and Norma, circa 1949.*

Bill married Norma Evans in 1941. She died in 1992. They had three sons, Vincent of Harrison, New York, Thomas of Sudbury, Massachusetts and Jody of New York City. Bill died in Chicago on April 22, 2005.

## ACKNOWLEDGMENT

We are indebted to Steve Koppes of the University of Chicago News Office who prepared an official obituary of Bill that was released by the University and from which we have borrowed extensively. The list of references that follows comprises a partial bibliography of William Kruskal. A full listing may be found at www.stat.uchicago.edu/people/inmemoriam.html.

## PARTIAL LIST OF PUBLICATIONS

[1] ADVISORY COMMITTEE ON PROBLEMS OF CENSUS ENUMERATION (1972). *America's Uncounted People.* National Academy of Sciences, Washington.

[2] BJERVE, P. J. and KRUSKAL, W. H. (1989). Kiaer, Anders Nicolai. *Encyclopedia of Statistical Sciences* **Suppl.** 78. Wiley, New York.

[3] CAMPBELL, D. T., KRUSKAL, W. H. and WALLACE, W. P. (1966). Seating aggregation as an index of attitude. *Sociometry* **29** 1–15. (Correction **30** 104.)

[4] CHAYES, F. and KRUSKAL, W. H. (1966). An approximate statistical test for correlations between proportions. *J. Geology* **74** 692–702. (Correction **78** 380.)

[5] DAVID, H. T. and KRUSKAL, W. H. (1956). The WAGR sequential *t*-test reaches a decision with probability one. *Ann. Math. Statist.* **27** 797–805. (Correction **29** 936.) MR0081045

[6] GOODMAN, L. A. and KRUSKAL, W. H. (1954). Measures of association for cross classifications. *J. Amer. Statist. Assoc.* **49** 732–764.

[7] GOODMAN, L. A. and KRUSKAL, W. H. (1959). Measures of association for cross classifications. II. Further discussion and references. *J. Amer. Statist. Assoc.* **54** 123–163.

[8] GOODMAN, L. A. and KRUSKAL, W. H. (1963). Measures of association for cross classifications. III. Approximate sampling theory. *J. Amer. Statist. Assoc.* **58** 310–364. MR0156400

[9] GOODMAN, L. A. and KRUSKAL, W. H. (1972). Measures of association for cross classifications. IV. Simplification of asymptotic variances. *J. Amer. Statist. Assoc.* **67** 415–421.

[10] GOODMAN, L. A. and KRUSKAL, W. H. (1974). Empirical evaluation of formal theory. *J. Math. Sociology* **3** 187–196.

[11] GOODMAN, L. A. and KRUSKAL, W. H. (1979). *Measures of Association for Cross Classifications.* Springer, New York. MR0553108

[12] KRUSKAL, W. H. (1946). Helmert's distribution. *Amer. Math. Monthly* **53** 435–438. MR0017490

[13] KRUSKAL, W. H. (1952). A nonparametric test for the several sample problem. *Ann. Math. Statist.* **23** 525–540. MR0050850

[14] KRUSKAL, W. H. (1953). On the uniqueness of the line of organic correlation. *Biometrics* **9** 47–58. MR0054224

[15] KRUSKAL, W. H. (1954). The monotonicity of the ratio of two noncentral *t* density functions. *Ann. Math. Statist.* **25** 162–165. MR0061320

[16] KRUSKAL, W. H. (1957). Historical notes on the Wilcoxon unpaired two-sample test. *J. Amer. Statist. Assoc.* **52** 356–360.

[17] KRUSKAL, W. H. (1958). Ordinal measures of association. *J. Amer. Statist. Assoc.* **53** 814–861. MR0100941

[18] KRUSKAL, W. H. (1960). Some remarks on wild observations. *Technometrics* **2** 1–3. MR0110572

[19] KRUSKAL, W. H. (1961). The coordinate-free approach to Gauss–Markov estimation and its application to

missing and extra observations. *Proc. Fourth Berkeley Symp. Math. Statist. Probab.* **1** 435–451. Univ. California Press, Berkeley. MR0137222

[20] KRUSKAL, W. H. (1965). Statistics, Molière and Henry Adams. *The Centennial Review* **9** 79–96.

[21] KRUSKAL, W. H., assoc. ed. for Statistics (1968). *International Encyclopedia of the Social Sciences.* Macmillan and Free Press, New York.

[22] KRUSKAL, W. H. (1968). When are Gauss–Markov and least squares estimators identical? A coordinate-free approach. *Ann. Math. Statist.* **39** 70–75. MR0222998

[23] KRUSKAL, W. H. (1971). Mathematical sciences and social sciences: Excerpts from the report of a panel of the Behavioral and Social Sciences Survey. *Amer. Statist.* **25**(1) 27–31.

[24] KRUSKAL, W. H. (1973). The Committee on National Statistics. *Science* **180** 1256–1258.

[25] KRUSKAL, W. H. (1974). The ubiquity of statistics. *Amer. Statist.* **28** 3–6.

[26] KRUSKAL, W. H. (1975). The geometry of generalized inverses. *J. Roy. Statist. Soc. Ser. B* **37** 272–283. (Correction **48** 258.) MR0379522

[27] KRUSKAL, W. H. (1978). Taking data seriously. In *Toward a Metric of Science*: *The Advent of Science Indicators* (Y. Elkana, J. Lederberg, R. K. Merton, A. Thackray and H. Zuckerman, eds.) 139–169. Wiley, New York.

[28] KRUSKAL, W. H. (1978). Leonard Jimmie Savage and Richard Price. Biographies in *International Encyclopedia of Statistics* (W. H. Kruskal and J. M. Tanur, eds.). Free Press, New York.

[29] KRUSKAL, W. H. (1978). Formulas, numbers, words: Statistics in prose. *The American Scholar* **47** 223–229.

[30] KRUSKAL, W. H. (1980). First interactions with Harold Hotelling; testing the Norden bombsight. *J. Amer. Statist. Assoc.* **75** 331–333.

[31] KRUSKAL, W. H. (1980). The significance of Fisher: A review of *R. A. Fisher. The Life of a Scientist*, by Joan Fisher Box. *J. Amer. Statist. Assoc.* **75** 1019–1030.

[32] KRUSKAL, W. H. (1982). Criteria for judging statistical graphics. *Utilitas Math.* **21B** 283–310. MR0683928

[33] KRUSKAL, W. H. (1984). The census as a national ceremony. In *Federal Statistics and National Needs* 177–180. U.S. Government Printing Office, Washington.

[34] KRUSKAL, W. H. (1986). Evaluating social science research. In *Advances in the Social Sciences, 1900–1980.* (K. W. Deutsch, A. S. Markovits and J. Platt, eds.) 231–247. Abt Books, Cambridge, MA.

[35] KRUSKAL, W. H. (1986). Terms of reference: Singular confusion about multiple causation. *J. Legal Studies* **15** 427–436.

[36] KRUSKAL, W. H. (1987). Relative importance by averaging over orderings. *Amer. Statist.* **41** 6–10. (Correction **41** 341.)

[37] KRUSKAL, W. H. (1988). The n cultures. In *Proc. Fourth Annual Research Conference* 231–236. Bureau of the Census, Washington.

[38] KRUSKAL, W. H. (1989). Hooker and Yule on relative importance: A statistical detective story. *Internat. Statist. Rev.* **57** 83–88.

[39] KRUSKAL, W. H. (1990). Contributions to "Fred as a scientific generalist." In *A Statistical Model*: *Frederick Mosteller's Contributions to Statistics, Science and Public Policy* (S. E. Fienberg, D. C. Hoaglin, W. H. Kruskal, J. M. Tanur and C. Youtz, eds.) 45–54. Springer, New York.

[40] KRUSKAL, W. H. (1997). Thermometers with detachable scales. Thomas Mann and the silent sisters of his Magic Mountain. In *Natur, Mathematik und Geschichte*: *Beiträge zur Alexander-von-Humboldt-Forschung und zur Mathematikhistoriographie* (H. Beck, R. Siegmund-Schultze, C. Suckow and M. Folkerts, eds.) 315–318. Deutsche Akademie der Naturforscher Leopoldina, Halle an der Saale.

[41] KRUSKAL, W. H. and MOSTELLER, F. (1979). Representative sampling. I. Non-scientific literature. *Internat. Statist. Rev.* **47** 13–24.

[42] KRUSKAL, W. H. and MOSTELLER, F. (1979). Representative sampling. II. Scientific literature, excluding statistics. *Internat. Statist. Rev.* **47** 111–127.

[43] KRUSKAL, W. H. and MOSTELLER, F. (1979). Representative sampling. III. The current statistical literature. *Internat. Statist. Rev.* **47** 245–265.

[44] KRUSKAL, W. H. and MOSTELLER, F. (1980). Representative sampling. IV. The history of the concept in statistics, 1895–1939. *Internat. Statist. Rev.* **48** 169–195. MR0586104

[45] KRUSKAL, W. H. and NEYMAN, J. (1995). Stochastic models and their applications to social phenomena. *Probab. Math. Statist.* **15** 21–27.

[46] KRUSKAL, W. H. and STIGLER, S. M. (1997). Normative terminology: "Normal" in statistics and elsewhere. In *Statistics and Public Policy* (B. Spencer, ed.) 85–111. Oxford Univ. Press.

[47] KRUSKAL, W. H. and TANUR, J. M., eds. (1978). *International Encyclopedia of Statistics.* Free Press, New York.

[48] KRUSKAL, W. H. and TELSER, L. G. (1960). Food prices and the Bureau of Labor Statistics. *J. Business* **33** 258–279, 285.

[49] KRUSKAL, W. H. and WALLIS, W. A. (1952). Use of ranks in one-criterion variance analysis. *J. Amer. Statist. Assoc.* **47** 583–621. (Correction **48** 907–911.)

[50] PRESIDENT'S COMMISSION ON FEDERAL STATISTICS (1971). *Federal Statistics: Report of the President's Commission.* U.S. Government Printing Office, Washington.

[51] TANUR, J. M., MOSTELLER, F., KRUSKAL, W. H., LINK, R. F., PIETERS, R. S. and RISING, G. R., eds. (1972). *Statistics*: *A Guide to the Unknown.* Holden-Day, San Francisco.

[52] ZABELL, S. (1994). A conversation with William Kruskal. *Statist. Sci.* **9** 285–303. MR1293298